\documentclass[twocolumn,showpacs,preprintnumbers,amsmath,amssymb, prl, reprint]{revtex4-1}
\usepackage{amsmath}
\usepackage{stmaryrd}
\usepackage{txfonts}
\usepackage{amssymb}
\usepackage{mathrsfs}
\usepackage{graphicx}
\usepackage{dcolumn}
\usepackage{bm}
\usepackage{epsfig}
\usepackage{color}
\usepackage{hyperref}

\begin{document}
\preprint{\href{http://dx.doi.org/10.1063/1.4861576}{S.-Z. Lin, C. Reichhardt, C. D. Batista and A. Saxena , J. Appl. Phys. {\bf 115}, 17D109 (2014).}}

\title{Dynamics of skyrmions in chiral magnets: dynamic phase transitions and equation of motion}
\author{Shi-Zeng Lin}
\affiliation{Theoretical Division, Los Alamos National Laboratory, Los Alamos, New Mexico
87545, USA}
\author{Charles Reichhardt}
\affiliation{Theoretical Division, Los Alamos National Laboratory, Los Alamos, New Mexico
87545, USA}
\author{Cristian D. Batista}
\affiliation{Theoretical Division, Los Alamos National Laboratory, Los Alamos, New Mexico
87545, USA}
\author{Avadh Saxena}
\affiliation{Theoretical Division, Los Alamos National Laboratory, Los Alamos, New Mexico
87545, USA}

\begin{abstract}
We study the dynamics of skyrmions in a metallic chiral magnet. First we show that skyrmions can be created dynamically by destabilizing the ferromagnetic background state through a spin polarized current. We then treat skyrmions as rigid particles and derive the corresponding equation of motion. The dynamics of skyrmions is dominated by the Magnus force, which accounts for the weak pinning of skyrmions observed in experiments. Finally we discuss the quantum motion of skyrmions. 

\end{abstract}
\date{\today}
\maketitle
Skyrmions as topological excitations were first proposed as a model for baryons by Skyrme half a century ago. \cite{Skyrme61}  Later on skyrmions were  realized in many different condensed matter systems, such as quantum Hall devices, multiband superconductors, liquid crystals and chiral magnets. The  observation of triangular skyrmion lattices in  metallic chiral magnets, like MnSi or $\mathrm{Fe_{0.5}Co_{0.5}Si}$, has sparked tremendous interest in this topological textures.~\cite{Muhlbauer2009,Yu2010a,Yu2011} For the case of  magnets, spins wrap a sphere once when moving from the center  to the outer region of the skyrmion. The typical size of an individual skyrmion and the skyrmion lattice constant is of order 10 - 100 nm. The skyrmion lattice was realized in a very small portion of the temperature-magnetic field phase diagram of bulk crystals , while it was found to be more stable for thin films and nanowires.~\cite{Yu2011,Heinze2011,Yu2013} Indeed,  skyrmions are found to be stable up to room temperature in FeGe.~\cite{Yu2011} More recently, skyrmions were also  discovered in insulating chiral magnets (e.g. $\mathrm{Cu_2OSeO_3}$)~\cite{Seki2012,Adams2012}. These findings suggest that skyrmions could be ubiquitous in magnetic materials without inversion symmetry.

In metallic magnets, the conduction electrons interact with skyrmions through the Hund's coupling, which is  larger than the Fermi energy. The spins of the conduction electrons are forced to be aligned with the local spins of the skyrmion, which yields emergent electromagnetic fields arising from the Berry phase that the electrons pick up because of the spin alignment.~\cite{Zang11} On the other hand, the skyrmion can also be driven by the electrons via the spin transfer torque mechanism. The measured threshold current density to make skyrmions mobile against pinning induced by defects is of order $10^6\ \mathrm{A/m^2}$, i.e., five orders of magnitude smaller than the depinning currents of magnetic domain walls.~\cite{Jonietz2010,Yu2012,Schulz2012} Consequently, skyrmions are believed to be promising for applications in spintronics, as they can be easily driven  by a spin polarized current.

For application purposes it is important to create skyrmions in a controlled fashion. The generation of skyrmions by electrical means would be advantageous to minimize device sizes. Because a skyrmion is a topological object, it is possible to treat it as a particle and to derive the corresponding equation of motion. The particle-like description of the skyrmion dynamics is more transparent and convenient for numerical simulations. In this paper, we summarize our recent study on skyrmion dynamics. First, we demonstrate a novel mechanism to create skyrmions by destabilizing the ferromagnetic (FM) state with a spin polarized current. Then, we present a particle-like equation of motion for skyrmions and explain the origin of the weak pinning. Finally, we study the quantum motion of skyrmions and discuss the possible experimental signature of the quantum effect of skyrmions.

We consider a thin film of a metallic chiral magnet described by the Hamiltonian \cite{Bogdanov89,Bogdanov94,Rosler2006,Han10,Rossler2011},
\begin{equation}\label{eq1}
\mathcal{H}=\int d\mathbf{r}^2 \left[\frac{J_{\rm{ex}}}{2}(\nabla \mathbf{n})^2+D\mathbf{n}\cdot\nabla\times \mathbf{n}-\mathbf{H}_a\cdot\mathbf{n} \right],
\end{equation} 
where $J_{\rm{ex}}$ is the exchange interaction, $D$ is the Dzyaloshinskii-Moriya (DM) interaction due to the absence of inversion symmetry in the system, and the last term is the Zeeman interaction with an external magnetic field $\mathbf{H}_a$. Here $\mathbf{n}$ is the unit vector representing the direction of the classical spin, and we have taken the continuum limit because the skyrmion size is much larger than the lattice constant. The system is assumed to be uniform along the $z$ direction. 

The phase diagram of $\mathcal{H}$ is known at $T=0$.~\cite{Rossler2011}   A single wave-vector magnetic spiral is stable at low fields. For intermediate fields, the triangular skyrmion lattice is stabilized  in order to lower the  Zeeman energy term (the magnetization along the field direction is nonzero for the skyrmion lattice solution). Finally, a fully polarized FM state becomes stable for large enough fields. The total number of skyrmions in the film is $Q=\int d\mathbf{r}^2q(\mathbf{r})$ with the skyrmion density $q(\mathbf{r}) = {\bf{n}}\cdot({\partial _x}{\bf{n}} \times {\partial _y}{\bf{n}})/(4\pi)$. \cite{SimonsQFT} The electric field generated by the skyrmion motion  is ${\bf{E}} ={\bf{n}}\cdot(\nabla {\bf{n}} \times {\partial _t}{\bf{n}})$. \cite{Zang11}. Dimensionless units are used throughout this paper.~\cite{szlin13skyrmion2}

\begin{figure}[t]
\psfig{figure=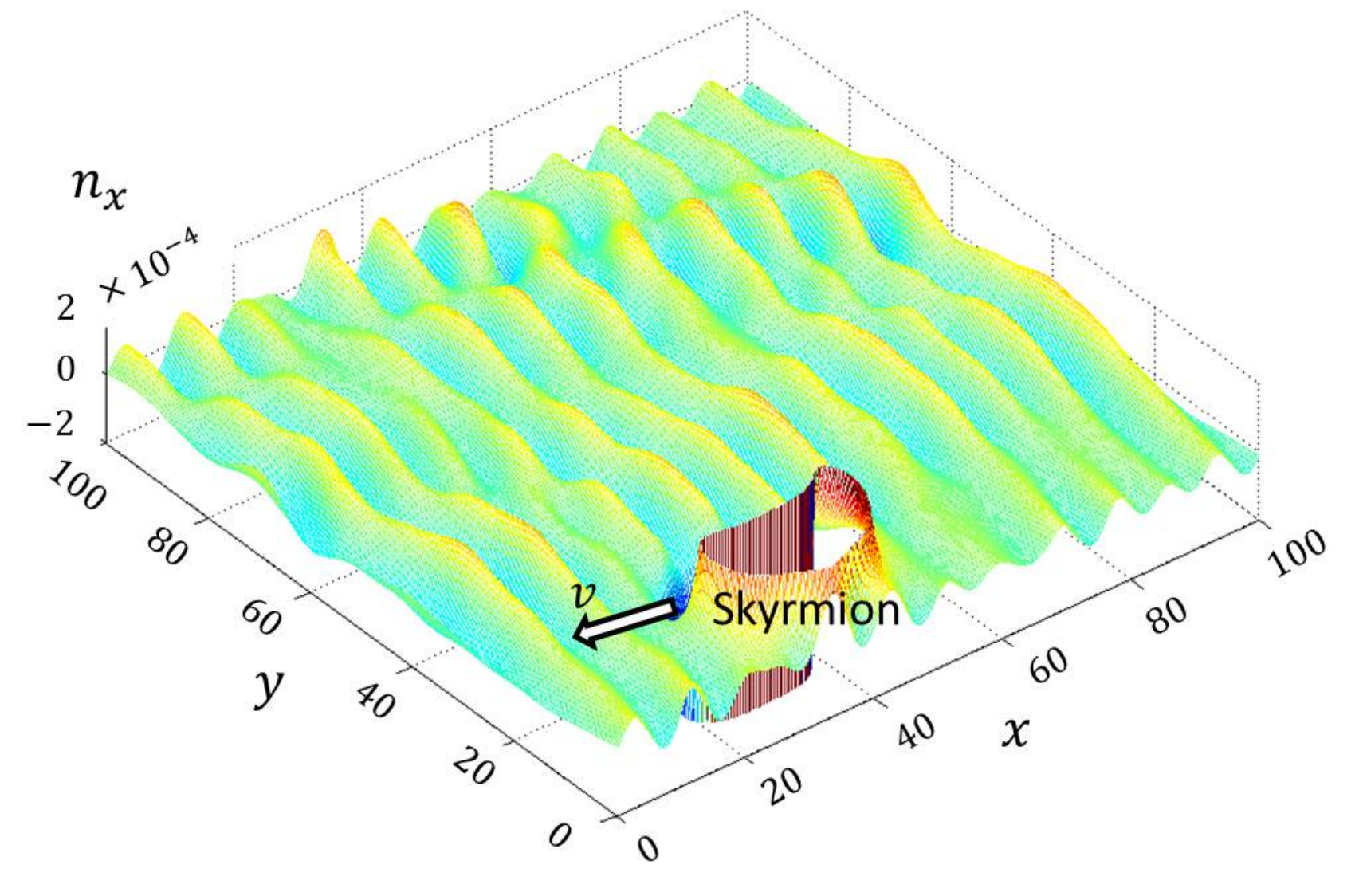,width=\columnwidth}
\caption{\label{f1}(color online) Spatial structure of the 
$n_x$ component of the spin wave radiated by the skyrmion motion. 
The skyrmion location is labelled in the figure and the arrow denotes the direction of skyrmion motion. Here $\alpha=0.1$, $J=1.4$ and $H_a=0.6$.}
\end{figure}
\begin{figure}[b]
\psfig{figure=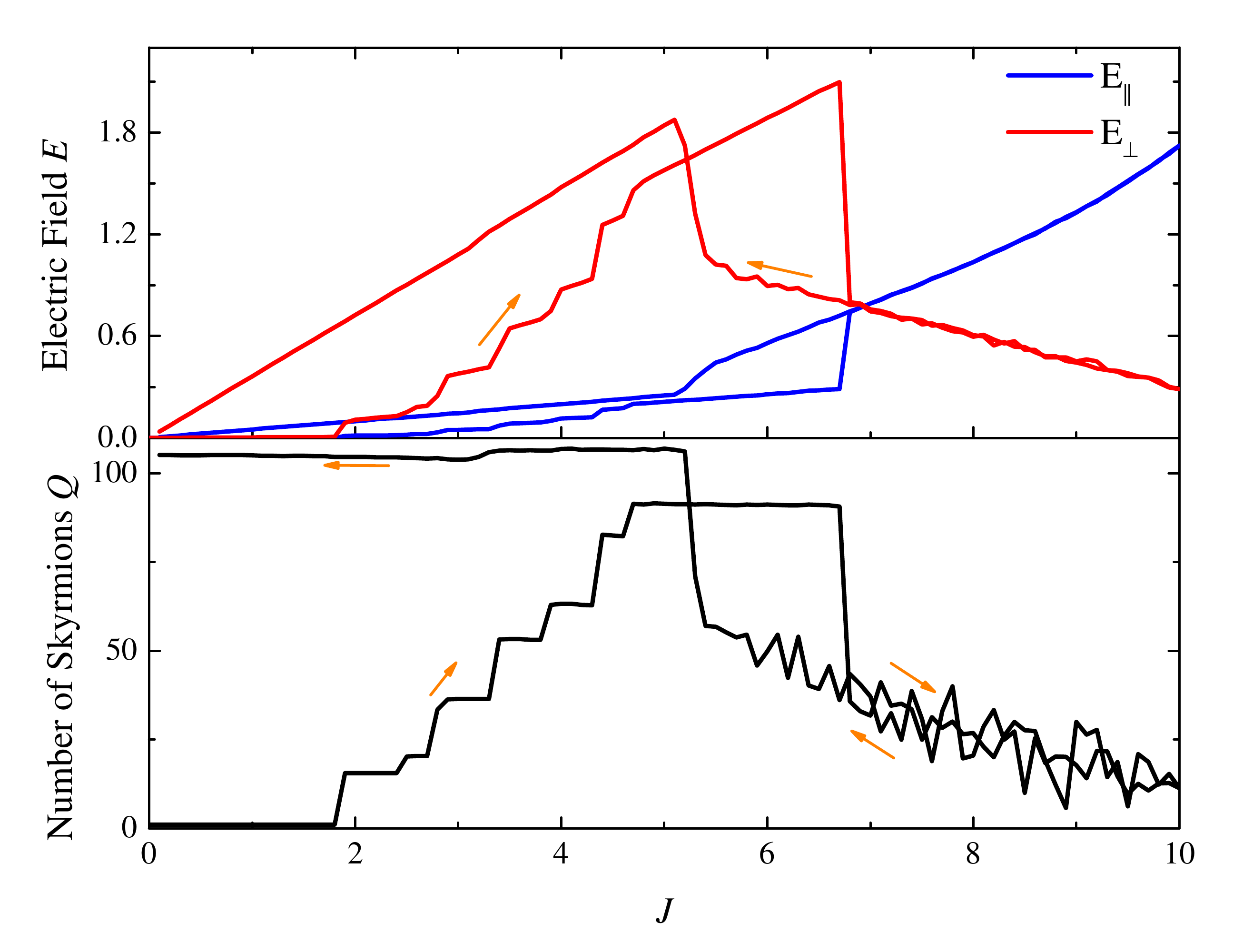,width=\columnwidth}
\caption{\label{f2}(color online) (a) Electric field parallel to the current $E_{\parallel}$ and the electric field perpendicular to the current $E_{\perp}$, (b) total number of skyrmions $Q$ as a function of the spin current $J$. The arrows in (a, b) indicate the direction of current sweep. Here $\alpha=0.1$ and $H_a=0.6$.}
\end{figure}

The spin dynamics  is described by the Landau-Lifshitz-Gilbert equation \cite{Bazaliy98,Li04,Tatara2008}
\begin{equation}\label{eq2}
{\partial _t}{\bf{n}} = ({{\bf{J}} }\cdot\nabla) {\bf{n}} - \gamma {\bf{n}} \times {{\bf{H}}_{\rm{eff}}} - \alpha {\partial _t}{\bf{n}} \times {\bf{n}},
\end{equation}
where the first term is the adiabatic spin transfer torque and the last term is the Gilbert phenomenological damping. We have $\alpha\ll 1$ for typical chiral magnets. Without current ($J=0$), the magnon dispersion in the FM state is $\Omega(\mathbf{J}=0, \mathbf{k})  = {\gamma (1+i\alpha) }\left(H_a+ J_{\rm{ex}} \mathbf{k}^2\right)/({\alpha ^2+1})$. The gap is induced by the applied magnetic field. However, $J$ can also be finite in presence of conduction electrons. If we take the conduction electrons as a reference frame, there is a Doppler shift of the magnon spectrum $\Omega(\mathbf{J}, \mathbf{k})=\Omega(\mathbf{J}=0, \mathbf{k})-\mathbf{k}\cdot \mathbf{v}$, where $\mathbf{v}=-\mathbf{J}$ is the velocity of the conduction electrons in dimensionless units. This Doppler shift was recently observed.~\cite{Vlaminck2008} The gap of the magnon spectrum vanishes for large enough current, $J_m={2 \gamma  \sqrt{H_a J_{\text{ex}}}}/({\alpha ^2+1})$, indicating that the FM state is no longer stable. In our simulations, we find that skyrmions are dynamically created  right after this instability~\cite{szlin13skyrmion1}  The reason is that the current density, $\mathbf{J}$, couples to the emergent vector potential $ \mathbf{A}\equiv {i c \hbar  b^{\dagger } \nabla b}/{e}$  generated by non-coplanar spin configurations via the Lagrangian term $\mathcal{L}_{JA}=\mathbf{J}\cdot\mathbf{A}$ (where $b$ is the spin coherent state \cite{SimonsQFT}). Spin states with non-zero $\mathbf{A}$ are favored by this coupling. In the presence of the DM interaction, the lowest energy state with non-zero $\mathbf{A}$ is a state with skyrmions . This mechanism can be used to create skyrmions by injecting current in a nanosized disk made of a chiral magnet.~\cite{szlin13skyrmion3}

To demonstrate the creation of skyrmions by current, we performed numerical simulations by putting a skyrmion in the FM state as an initial condition. The skyrmion radiates spin waves when it is driven by a current,  as shown in Fig.~\ref{f1}. The spin gap vanishes at a threshold current   and more skyrmions are dynamically created. The number of skyrmions increases continuously after the instability and finally saturates [see Fig.~\ref{f2} (b)]. Meanwhile, as it is shown in Fig. \ref{f2} (a), the electric field increases because it is proportional to the density of skyrmions. However, the ratio of the Hall electric field $E_{\perp}$ to the longitudinal electric field $E_{\parallel}$ is independent of $\mathbf{J}$ because it is an intrinsic property of rigid skyrmions.  This threshold current obtained from our simulations is consistent with the analytical result for $J_m$. For stronger drives, the skyrmions become strongly distorted (they are no longer circular objects) because the spin precession cannot follow the fast moving skyrmion. As a result, skyrmions are destroyed at such high current densities via the softening of some of its internal modes. For typical parameters, the current densities for generating and destroying skyrmions are the order of $10^{12}\ \mathrm{A/m^2}$.~\cite{szlin13skyrmion1}

\begin{figure}[t]
\psfig{figure=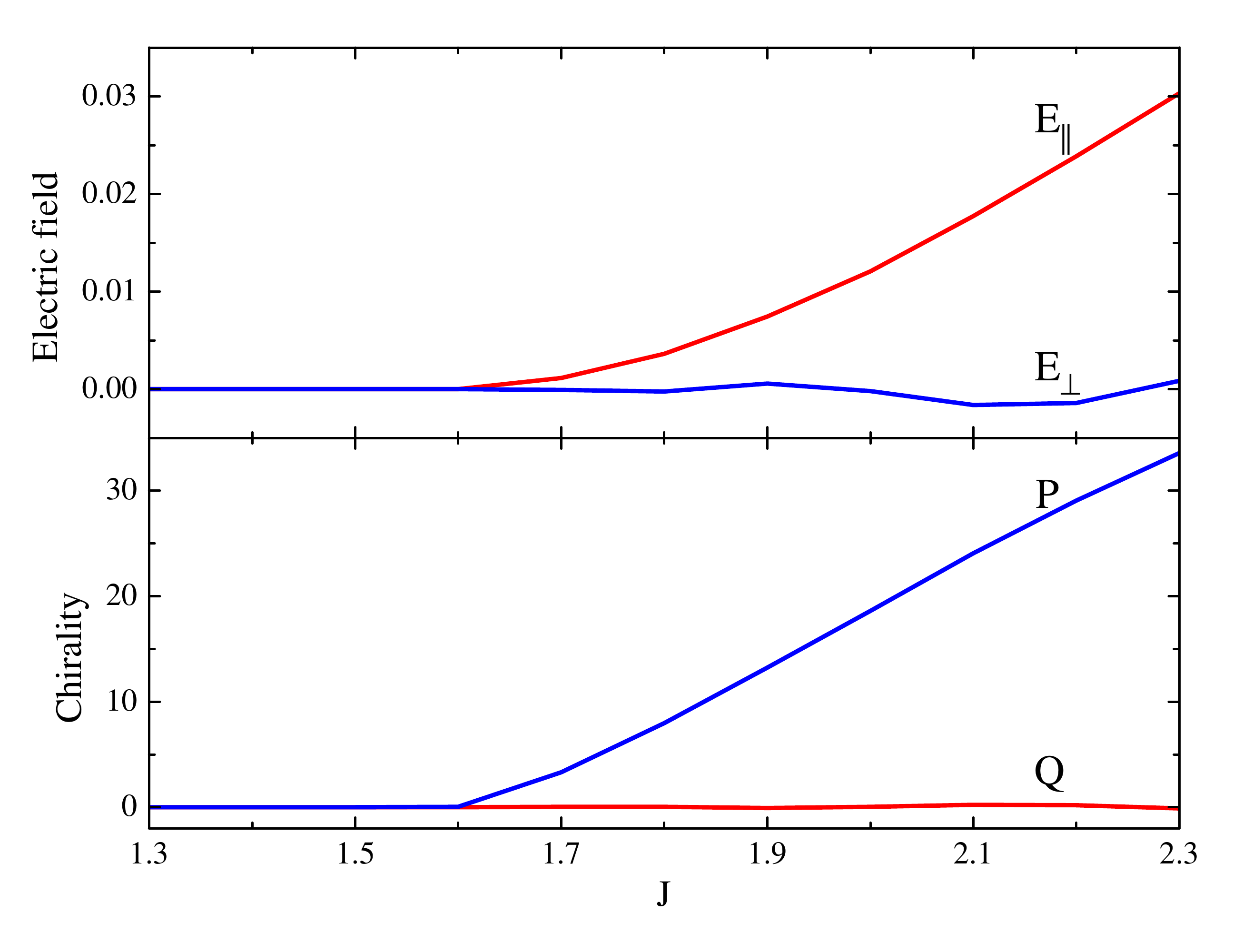,width=\columnwidth}
\caption{\label{f3}(color online) (a) Electric field parallel to the current, $E_{\parallel}$, and the electric field perpendicular to the current, $E_{\perp}$; (b) total number of skyrmions $Q$ and absolute chirality $P$ as a function of the spin current, $J$, when $J$ is increased.  Here $\alpha=0.1$, $D=0$ and $H_a=0.6$.}
\end{figure}

The current-induced instability of the FM state  also occurs for $D=0$. Because the current density is coupled with the vector potential in the Lagrangian, a states with a nonzero $\mathbf{A}$ is created to minimize the energy after the instability, similar to the case when $D\neq 0$ we explained above. However, there is no preferred chirality  for $D=0$. Consequently, a chiral liquid phase is stabilized according to our simulations. To characterize the chiral liquid phase, we introduce the averaged chirality $Q$ and the absolute chirality $P\equiv\int d\mathbf{r}^2|q(\mathbf{r})|$. As shown in Fig. \ref{f3}, $Q\approx 0$ while $P>0$  fluctuates strongly over space and time, indicating a chiral liquid phase. A nonzero electric field is also generated by the fluctuations of chirality [see Fig. \ref{f3}(a)]. Note that in a metal, the dominant contribution to the electric field comes from electrons. 

\begin{figure}[t]
\psfig{figure=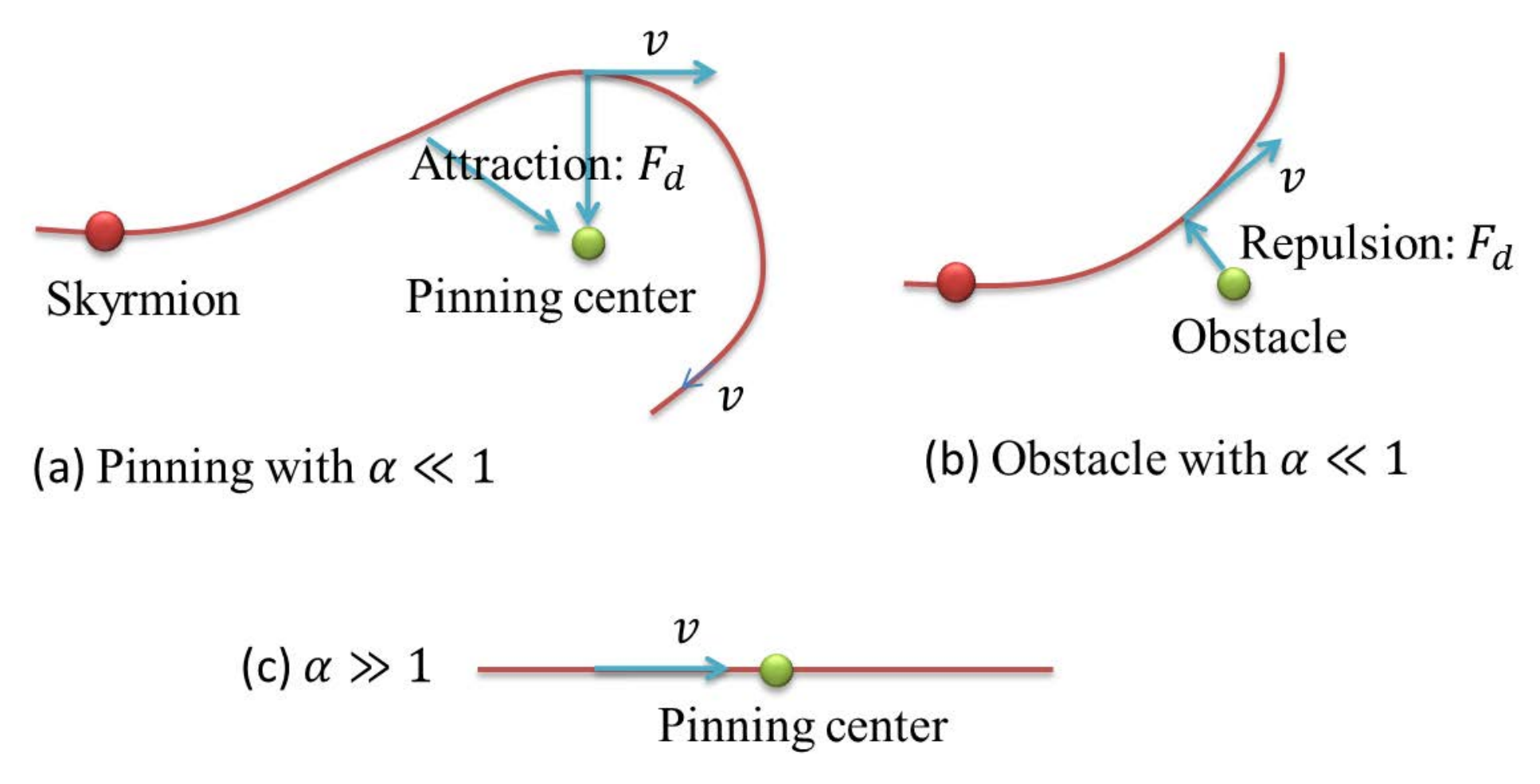,width=\columnwidth}
\caption{\label{f4}(color online) (a) and (b): Schematic view of a skyrmion passing through a pinning center (a) and obstacle (b) when the Magnus force is dominant. When the Magnus force is dominant over the dissipative force, the skyrmion is deflected by the pinning centers or obstacles. (c) Same as (a) and (b) except that the dissipative force is dominant. The skyrmion has to overcome the pinning site or obstacle by passing through it.}
\end{figure}

Skyrmions are topological excitations and can be treated as particles if the internal modes are not excited. In this case, their equation of motion is \cite{szlin13skyrmion2}
\begin{equation}\label{eq3}
\frac{4\pi\alpha  }{\gamma}\mathbf{v}_i=\mathbf{F}_M+\mathbf{F}_L+\sum_j\mathbf{F}_d(\mathbf{r}_j-\mathbf{r}_i)+\sum_j \mathbf{F}_{ss}(\mathbf{r}_j-\mathbf{r}_i),
\end{equation} 
which can be derived by using Thiele's approach \cite{Thiele72}. The term on the left-hand side accounts for the damping of skyrmion motion, which is produced by the underlying damping of the spin precession and damping due to conduction electrons that are localized around the skyrmions (for metals).  $\mathbf{F}_M=4\pi\gamma^{-1}\hat{z}\times \mathbf{v}_i$ is the Magnus force per unit length, which is perpendicular to the velocity. $\mathbf{F}_L=2\pi\hbar e^{-1}\hat{z}\times \mathbf{J}$ is the Lorentz force due to the external current, which arises from the emergent quantized magnetic flux $\Phi_0=hc/e$ carried by the skyrmion in the presence of a finite current.  $\mathbf{F}_d$ is the interaction between skyrmions and quenched disorder and $\mathbf{F}_{ss}$ is the short-range pairwise interaction between two skyrmions. The damping is weak $\alpha\ll 1$ and the Magnus force $\mathbf{F}_M$ is dominant over the dissipative force ${4\pi\alpha  }\mathbf{v}_i/\gamma$. Equation~\eqref{eq3} also describes the skyrmion motion in insulators, where the dissipation due to conduction electrons is absent and $\mathbf{F}_L=0$.

The rather strong Magnus is one reason behind the weak pinning that has been observed for skyrmions. In the presence of a pinning center or an obstacle, skyrmions can easily be scattered with a velocity perpendicular to the pinning  or repulsive force. Thus, as illustrated in Fig.~\ref{f4} (a) and (b), the influence of the pinning center or obstacle can be minimized by avoiding passing through them. When the dissipative force is dominant, $\alpha \gg 1$,  the pinning force becomes very strong because skyrmions have to pass through the pinning center, as sketched in Fig.~\ref{f4}(c). A similar conclusion was also reached by numerical simulations of the continuum model [Eqs.~\eqref{eq1} and \eqref{eq2}].~\cite{Iwasaki2013}

Finally, we discuss the quantum motion of skyrmions by quantizing Eq.~\eqref{eq3}.~\cite{szlin13skyrmion4} In a clean sample, skyrmions occupy the lowest Landau level with their wave function strongly localized due to the strong emergent magnetic field. In the presence of a pinning potential, the lowest Landau level for skyrmions is split into quantized levels. The transition between different levels can be observed experimentally by microwave absorption measurements in the frequency region around $\omega\approx 100$ GHz  and at low temperatures $T<0.5$ K for typical materials parameters.

\noindent \textit{Acknowledgements --} Computer resources for numerical calculations were provided by the Institutional Computing Program in LANL. This publication was made possible by funding from the Los Alamos Laboratory Directed Research and Development Program, project number 20110181ER.

%

\end{document}